\documentclass[12pt]{iopart}

\usepackage{iopams}  
\usepackage{graphicx}
\begin{document}

\title[Mathematical analysis of unstable density fluctuations]{Mathematical analysis of unstable density fluctuations in 
the dissipative
gravitational collapse}

\author{A. R. M\'endez}
\address{Depto. de Matem\'aticas Aplicadas y Sistemas, Universidad Aut\'onoma Metropolitana-Cuajimalpa, Prol. Vasco de Quiroga 4871, 
M\'exico D.F 05348, M\'exico.}
\ead{amendez@correo.cua.uam.mx}

\author{A. L. Garc\'ia-Perciante}
\address{Depto. de Matem\'aticas Aplicadas y Sistemas, Universidad Aut\'onoma Metropolitana-Cuajimalpa, Prol. Vasco de Quiroga 4871, 
M\'exico D.F 05348, M\'exico.}
\ead{algarcia@correo.cua.uam.mx}

\author{D. M. Ruiz-Moreno}
\address{Depto. de F\'isica y Matem\'aticas, Universidad Iberoamericana, Prolongaci\'on Paseo de la Reforma 880, M\'exico D. F. 01219, M\'exico.}

\author{A. Sandoval-Villalbazo}
\address{Depto. de F\'isica y Matem\'aticas, Universidad Iberoamericana, Prolongaci\'on Paseo de la Reforma 880, M\'exico D. F. 01219, M\'exico.}
\ead{alfredo.sandoval@ibero.mx}

\vspace{10pt}
\begin{indented}
\item[]July 2018.
\end{indented}

\begin{abstract}
A detailed analysis of the dynamics of unstable modes present in the
linearized Navier-Stokes-Fourier system in the presence of a gravitational field is carried out. The transition
between the non-dissipative and dissipative regimes is explored from
the mathematical point of view by considering the same equation of
state in both cases, a procedure which sheds light on the discontinuity
present in the critical parameter. It is also shown that two distinct
behaviours for the propagation of unstable modes can be identified
below the critical value for a gravitational instability to occur,
depending on a parameter $r$ that quantifies the relevance of dissipation
relative to gravitational effects.
\end{abstract}

%
%
%
%
%

\section{Introduction}
Structure formation is a topic of great interest, which has been addressed
by several authors in the last decades \cite{CoronaGalindo,Carlevaro, GColin-Sandoval2002,GColin-Sandoval2005,Kumar, Archana,Mendez-Garcia2016}. In very
general terms, the model for such a process consists in considering
a homogeneous gas in which a spontaneous fluctuation in density
locally gives rise to a gravitational field. As the
density fluctuations increase, the balance between gas pressure
and gravitational potential gradient can be broken for fluctuations
with a wave number below a critical value, namely the Jeans wave number
\cite{Garcia-Mendez-Sandoval,Jeans1902}. When the corresponding criterion is met, a gravitational
collapse is initiated, leading to structure formation. 

In his original work, Jeans considered an ideal non-dissipative system
for which the linearized transport equations for fluctuations can
be written as
\begin{equation}
\frac{\partial\left(\delta n\right)}{\partial t}+n_{0}\delta\theta=0,\label{1}
\end{equation}

\begin{equation}
\frac{\partial\left(\delta\theta\right)}{\partial t}+\frac{1}{mn_{0}}\nabla^{2}\left(\delta p\right)=-4\pi Gm\delta n,\label{2}
\end{equation}

\begin{equation}
C_{n}n_{0}\frac{\partial\left(\delta T\right)}{\partial t}+p_{0}\delta\theta=0,\label{3}
\end{equation}
where $\theta$ is the divergence of the hydrodynamic velocity, $n$
the local particle density, $m$ the mass of the molecules, $p$ the
hydrostatic pressure, $T$ the temperature, $G$ the gravitational
constant and $C_{n}$ the specific heat at constant density. The state
variables are considered to have a constant, equilibrium, value and
a fluctuating component as follows
\begin{equation}
X=X_{0}+\delta X,\label{4}
\end{equation}
The stability of the system of equations above can be analyzed by
direct algebraic manipulation in Fourier-Laplace space in order to
yield the following dispersion relation:
\begin{equation}
s^{2}=1-\frac{q^{2}}{q_{J}^{2}}.\label{5}
\end{equation}
In Eq. (\ref{5}), $s$ and $q$ are the corresponding
Fourier and Laplace transform parameters, and $q_{J}^{2}=4\pi Gmn_{0}/C_{s}^{2}$
is the Jeans wavenumber with $C_{s}$ being the adiabatic speed of
sound and $G$ the gravitational constant. The relation given by Eq.
(\ref{5}) implies that the system is stable for $q^{2}>q_{J}^{2}$.
Such criterion has been widely employed to roughly determine the mass
required in a molecular cloud for a gravitational collapse to be feasible. 

Even though the criterion described above serves as a good approximation,
other factors may in principle enhance or hinder the growth of fluctuations,
such as dissipation and external fields. The analysis of the effect
that such phenomena have in the Jeans problem is not straightforward
and has been studied to certain extent \cite{CoronaGalindo,Carlevaro, 
GColin-Sandoval2002,GColin-Sandoval2005,Kumar, Archana,Mendez-Garcia2016}. Indeed,
when dissipation is taken into account, the dispersion relation leads
to a complete cubic polynomial and thus a more involved analysis is required
in order to establish the threshold for exponential growth of fluctuations
as well as to explore their dynamics. Several attempts have been made
in order to account for thermal and viscous dissipation in the onset
of the collapse using different approaches. Kumar showed, applying
the Routh Hurwitz criterion, that viscous dissipation, rotation and
the presence of a magnetic field do not affect the Jeans number in
the case of an ionized gas \cite{GColin-Sandoval2005}. However, thermal dissipation
was shown to alter the criterion in a factor, which does not depend
on thermal conductivity, as follows
\begin{equation}
q_{JD}^{2}=\gamma\frac{4\pi Gmn_{0}}{C_{s}^{2}},\label{6}
\end{equation}
where $\gamma$ is the adiabatic index \cite{GColin-Sandoval2002,Archana}. Clearly, this
jump in the parameter represents an abrupt change in the behavior
of the solutions to the system and does not reduce exactly to the
non-dissipative value when the corresponding transport coefficient
is taken to be zero. On the other hand, alternative methods have also
been used in order to approximate the solutions to the dispersion
relation and to study the corresponding threshold and dynamics of
fluctuations. In particular, in Refs. \cite{GColin-Sandoval2002,GColin-Sandoval2005} a factorization
for the dispersion relation is proposed, however the threshold in
those cases does depend on transport coefficients, opposite to the
criterion obtained analytically in Ref. \cite{Kumar}.

The aim of the present paper is to revisit the study of density fluctuation
dynamics in the presence of dissipation for a self-gravitating system
in a thorough and mostly analytical fashion. In particular, the transition
between the Jeans wave numbers corresponding to the ideal and dissipative
scenarios is analyzed. Also, two specific features of the dynamics of fluctuations below the
critical parameter are studied and two regimes are identified.

In order to establish and describe these results, the rest of this
work is organized as follows. In Sect. 2, the linearized transport
equations within the Navier-Stokes-Fourier regime are shown and an
dimensionless dispersion relation is obtained. The Routh-Hurwitz criterion
is applied in order to establish the modified criterion for the dissipative
gas. Two distinct behaviors for unstable modes are identified below
the critical value in Sect. 3. Section 4 is devoted to the analysis
of the scenario which includes the low dissipation case and the transition
to the result obtained without dissipation is addressed. A thorough
discussion of the results and final remarks are included in Sect.
5.

\section{Transport equations and Routh-Hurwitz stability analysis}

For a self-gravitating system in the linearized Navier-Stokes-Fourier
regime, the set of hydrodynamic equations read

\begin{equation}
\frac{\partial\left(\delta n\right)}{\partial t}+n_{0}\delta\theta=0,\label{7}
\end{equation}

\begin{equation}
\frac{\partial\left(\delta\theta\right)}{\partial t}+\frac{p_{0}}{\rho_{0}}\left(\frac{\nabla^{2}\left(\delta T\right)}{T_{0}}+\frac{\nabla^{2}\left(\delta n\right)}{n_{0}}\right)-\frac{4}{3}\frac{\eta}{mn_{0}}\nabla^{2}\left(\delta\theta\right)=-4\pi Gm\delta n,\label{8}
\end{equation}

\begin{equation}
C_{n}n_{0}\frac{\partial\left(\delta T\right)}{\partial t}+p_{0}\delta\theta-\frac{\kappa}{T_{0}}\nabla^{2}\left(\delta T\right)=0,\label{9}
\end{equation}

\noindent where $\eta$ is the shear viscosity and $\kappa$ the thermal
conductivity. For the sake of simplicity, a monoatomic ideal gas is
here considered and the transport coefficients are expressed in terms
of a relaxation time $\tau$, following the BGK method in Boltzmann's
equation \cite{Chapman}. That is,
\begin{equation}
\eta=nkT\tau,\qquad \kappa=\frac{5}{3}nkT\tau,\qquad C_{s}^{2}=\frac{5}{3}\frac{kT}{m}.\label{10}
\end{equation}
In Fourier-Laplace space, the system (\ref{7}-\ref{9}) can be shown
to lead to a third order polynomial dispersion relation which can
be expressed as
\begin{equation}
a_{3}S^{3}+a_{2}S^{2}+a_{1}S+a_{0}=0,\label{11}
\end{equation}
where the coefficients are given by
\begin{eqnarray}\nonumber
 a_{3}&=1,\\\nonumber
 a_{2}&=\frac{9}{5}KR,\\\nonumber
 a_{1}&=\frac{4}{5}R^{2}K^{2}+K-1,\\\nonumber
 a_{0}&=\left(\frac{3}{5}K-1\right)RK.
\end{eqnarray}
Here, the normalized wavenumber and frequency are given by $K=\left(q/q_{J}\right)^{2}$
and $S=s\tau_{G}$ respectively, where $\tau_{G}=\left(4\pi Gmn_{0}\right)^{-1/2}$
corresponds to a gravitational characteristic time. The dissipation
parameter is here defined as $R=\tau/\tau_{G}$ and compares the typical
collisional (microscopic) time to the characteristic gravitational
timescale. 

The Routh-Hurwitz criterion can be readily applied to Eq. (\ref{11}),
by means of which the number of roots on the right side quadrants
of the complex plain correspond to the number of sign changes in the
array $\left\{ a_{3},\,a_{2},\,b_{1},\,a_{0}\right\} $ with
\[
b_{1}=\frac{a_{2}a_{1}-a_{3}a_{0}}{a_{2}}.
\]
Stability requires that all roots of Eq. (\ref{11}) have negative
real parts, such that fluctuations decay in time. Since $a_{3}$ and
$a_{2}$ are positive, the condition for a stable system is given
by
\[
0<a_{0}<a_{2}a_{1}.
\]
This condition is satisfied for $R\neq0$, 
\begin{equation}
K>\frac{5}{3},\label{12}
\end{equation}
and
\begin{equation}
\frac{6}{5}R^{2}K^{2}+K-\frac{2}{3}>0,\label{13}
\end{equation}
simultaneously. Clearly, Eq. (\ref{13}) is always satisfied as long
as Eq. (\ref{12}) holds true and thus the criterion for stability
is given by $K>5/3$, which is independent of the particular value
of $R$ provided that $R\neq0$. That is, if $R>0$
and $K>5/3$ the three roots of Eq. (\ref{11}) have negative real
parts which yields an exponential decay of fluctuations in time and
thus a stable system.

\begin{figure}
\begin{center}
\includegraphics[scale=0.5]{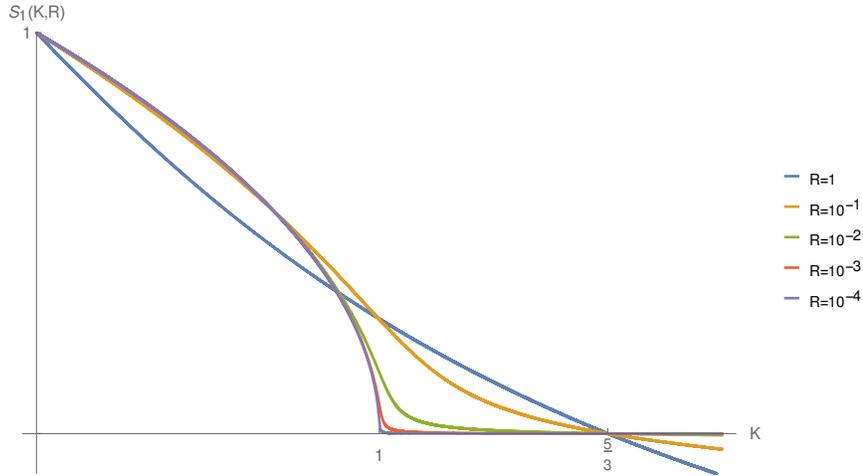}
\caption{\label{fig:1} The real root $S_{1}\left(K,R\right)$ when $0<K<5/3$, as a function of the dimensionless wavenumber $K$ for different values of $R$.}
\end{center}
\end{figure}

On the other hand for $0<K<5/3$ one has $a_{0}<0$ which indicates
only one sign change in $\left\{ a_{3},\,a_{2},\,b_{1},\,a_{0}\right\} $.
This means that one, and only one of the three roots lies in the right
side of the complex plane and is thus real. Let $S_{1}\left(K,R\right)$
denote such root. Figure \ref{fig:1} shows a plot of $S_{1}\left(K,R\right)$
as a function of $K$ for several values of $R$ where it can be seen
that this root is always real and changes sign in $K=5/3$. Also notice
that for $R\rightarrow0$, $S_{1}$ approaches zero rapidly around
$K=1$, having an almost vertical tangent. However, as will be discussed
in Sect. 4, $S_{1}\left(K,R\right)=0$ only for $K=5/3$ if $R\neq0$.
The nature of the remaining roots depends strongly on the value of
the dissipation parameter $R$ and is analyzed in the next section.
\section{Dynamics of unstable modes}

One of the three roots of the dispersion relation is real for all
values of $K$, independently of the parameter $R$ (as long as $R>0$).
In this section, the behavior of the two remaining roots, call them $S_2$ and $S_3$, is analyzed
within the unstable region in Laplace space, i. e. $0\leq K<5/3$.
It has already been shown, based on the Routh Hurwitz criterion, that
the real part of $S_{2}$ and $S_{3}$ is negative, corresponding
to stable modes. However, the details of the propagation, that is
if fluctuations oscillate or are purely damped, depends on whether
these roots are complex or real. This is determined
by the sign of the discriminant of the dispersion relation
\[
d=a_{1}^{2}a_{2}^{2}-4a_{0}a_{2}^{3}-4a_{1}^{3}a_{3}+18a_{0}a_{1}a_{2}a_{3}-27a_{0}^{2}a_{3}^{2},
\]
which is a sixth order polynomial both in $K$ and $R$ and thus its
analysis is not straightforward. However, since it features only even
powers of $R$, we define $r=R^{2}$ in order to simplify the notation. The condition 
for the presence of complex roots is written as $\Delta\left(K,r\right)<0$
where $\Delta$ is defined as
\begin{eqnarray}
\Delta\left(K,r\right) & =\frac{4}{625}\left(4r^{3}K^{6}-147r^{2}K^{5}+\left(525r-15r^{2}\right)K^{4}-\left(1050r+625\right)K^{3}\right.\label{eq:desigualdad-1}\\
 & \left.-\left(150r-1875\right)K^{2}-1875K+625\right).\nonumber 
\end{eqnarray}
Since $\Delta\left(0,\,r\right)>0$, one has three real roots with
single multiplicity when $K=0$. However, $S_{2}$ and $S_{3}$ may
become complex if $\Delta\left(K,r\right)$ becomes negative at some
$K$, for a particular value of $r$. Since
\[
\Delta\left(\frac{5}{3},r\right)=\left(\frac{2}{27}\left(3+10r\right)\right)^{2}\left(r-24\right),
\]
it is clear that for $r<24$ the discriminant changes
sign at least once within the unstable region. Figure \ref{fig:2} shows
that for $\left(K,r\right)\in\left[0,5/3\right]\times\left[0,24\right]$,
$\Delta$ changes sign only once. 
\begin{center}
\begin{figure}
\begin{center}
\includegraphics[scale=0.7]{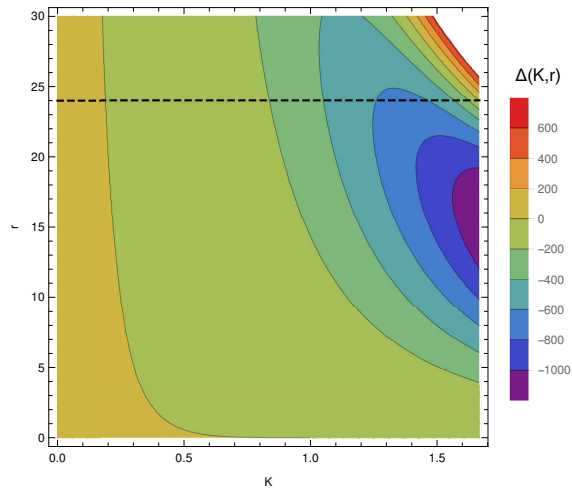}
\end{center}
\caption{\label{fig:2} Level curves of the discriminant
given in (\ref{eq:desigualdad-1}).}
\end{figure}
\par\end{center}

The change of sign in the discriminant indicates that there is a critical
wavenumber for which the stable modes become complex and thus would
lead to a doublet in a scattering experiment spectrum \cite{Berne}.
Moreover, once these modes change in nature, their damping weakens
and thus the characteristic damping time increases. Figure \ref{fig:3}
shows the different behaviors of the three modes for $R=0.1$. This
behavior corresponds to a low dissipation case, in which the characteristic
form of a Rayleigh-Brillouin spectrum is obtained for $K>5/3$, and
the Brillouin doublet is present even within the unstable region.
\noindent \begin{center}
\begin{figure}
\begin{center}
\includegraphics[scale=0.85]{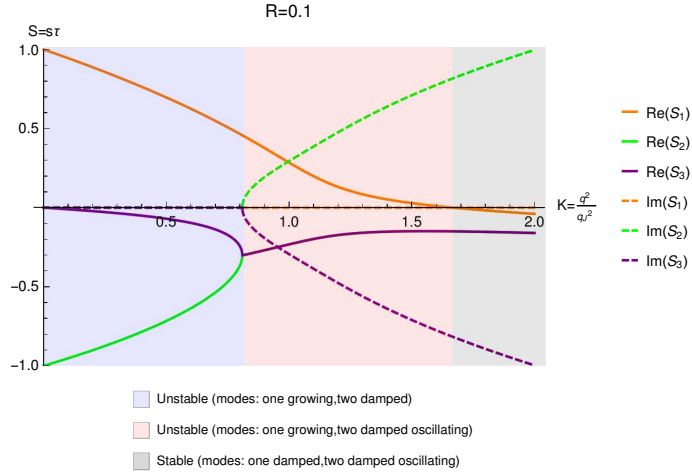}
\end{center}
\caption{\label{fig:3}The real (continuous line) and imaginary (dashed line)
parts of the roots of the dispersion relation for $R=0.1$. }
\end{figure}
\par\end{center}

\begin{figure}
\begin{centering}
\includegraphics[scale=0.5]{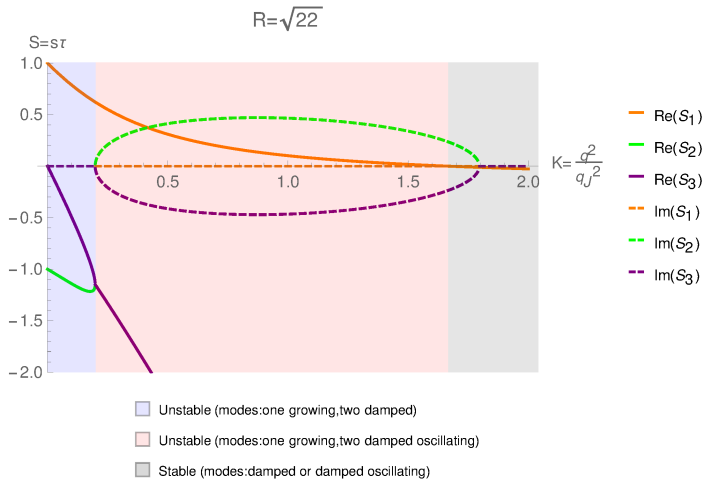}
\par\end{centering}
\caption{\label{fig:4}The real (continuous line) and imaginary (dashed line)
parts of the roots of the dispersion relation for $r<24$ ($R=\sqrt{22})$. }
\end{figure}

Figure \ref{fig:4} shows that there is a second root, and change
of sign in $\Delta$, into the stable region which
approaches the critical value $K=5/3$ as $r\rightarrow24$. The $r=24$
limiting case is illustrated in Fig. \ref{fig:5} where the change
in behavior of the conjugate roots occurs simultaneously with the
onset of stability. In that case one expects to obtain a finite spectrum
for $K>5/3$ featuring only a superposition of three central peaks
with no Brillouin doublet. For $r>24$ (see Fig. \ref{fig:6}) the
doublet also disappears within the unstable region and could not be
observed for wavenumbers above $5/3$. 
\begin{center}
\begin{figure}
\begin{centering}
\includegraphics[scale=0.5]{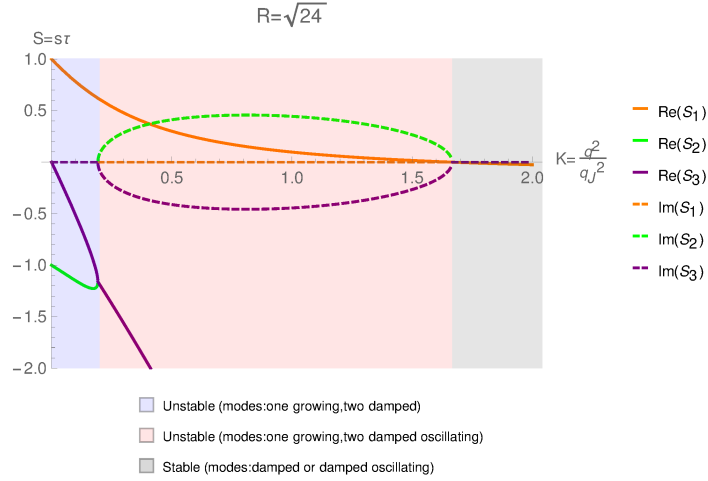}
\par\end{centering}
\caption{\label{fig:5} The real (continuous line) and imaginary (dashed line)
parts of the roots of the dispersion relation for the limiting case
$r=24$ (or $R=\sqrt{24}$). }
\end{figure}
\par\end{center}

\begin{center}
\begin{figure}
\begin{centering}
\includegraphics[scale=0.5]{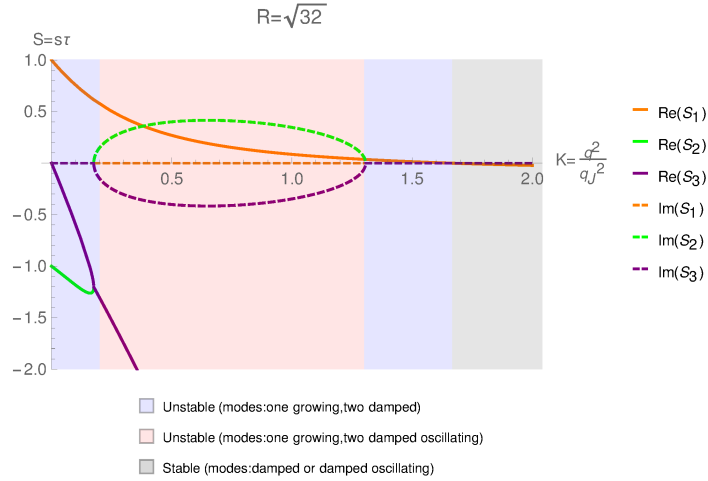}
\par\end{centering}
\caption{\label{fig:6} When $r>24$ the Brillouin-doublet disappears within
the unstable region and is not observed for wavenumbers over $5/3$.}
\end{figure}
\par\end{center}
\section{Asymptotic behavior for low dissipation}

\noindent The analysis carried out in the previous sections leads
naturally to the question on whether the non-dissipative limit is
obtained when $R\rightarrow0$. That is, as strongly emphasized in
Section 1, the critical wavenumber for structure formation is given
by $K=5/3$ for $R\neq0$. However, if $R=0$ one finds the reported
value in the literature corresponds to $K=1$ \cite{Jeans1928}.
Moreover, in the non-dissipative case the dispersion relation is
a quadratic polynomial which has two real roots for $K<1$ and two
imaginary roots if $K>1$ (see fig. \ref{fig:7}). 
\begin{center}
\begin{figure}
\begin{centering}
\includegraphics[scale=0.32]{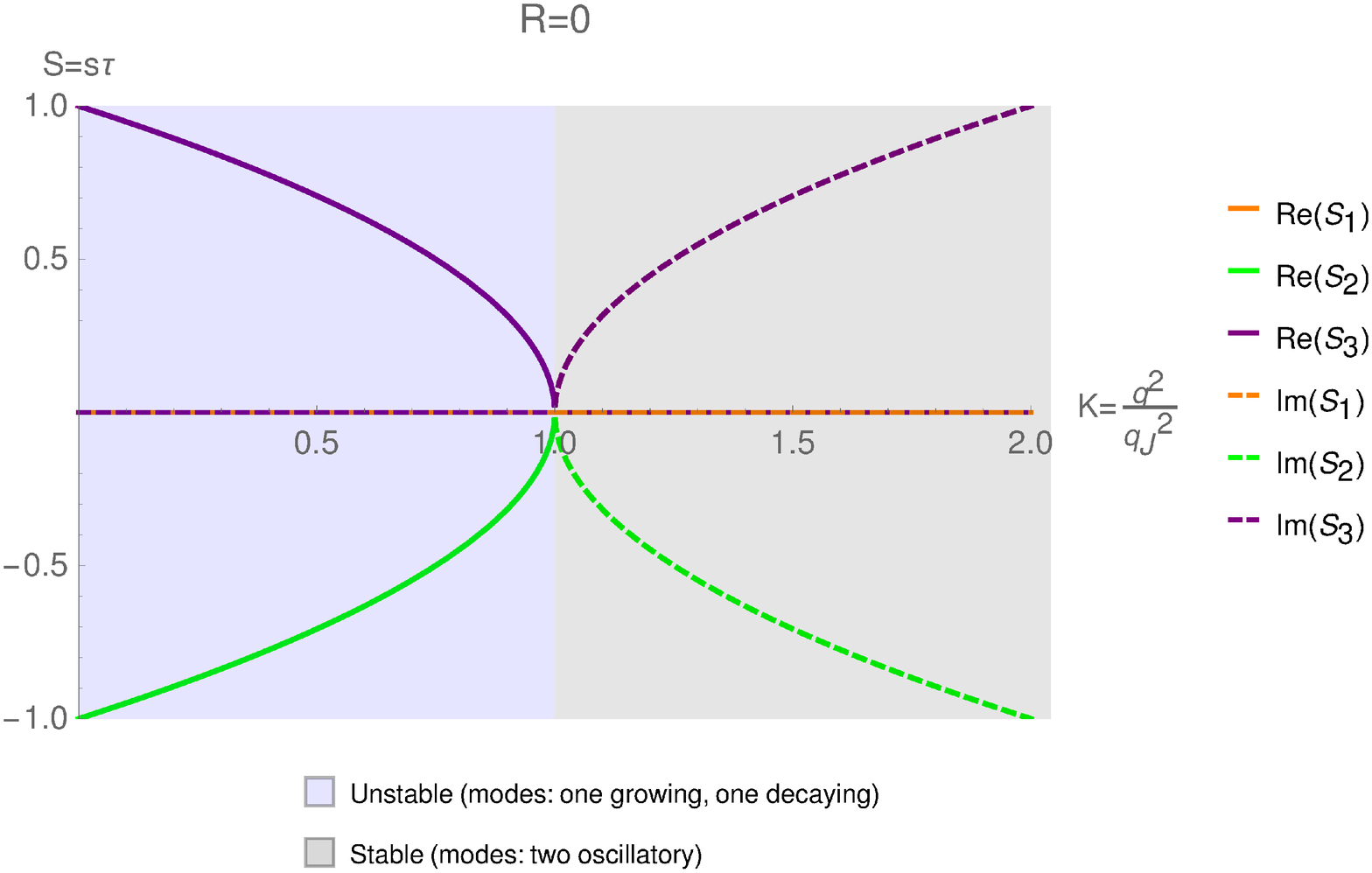}
\par\end{centering}
\caption{\label{fig:7}The real (continuous line) and imaginary (dashed line)
parts of the roots of the non-dissipative ($R=0$) dispersion relation. }
\end{figure}
\par\end{center}

\begin{center}
\begin{figure}
\begin{centering}
\includegraphics[scale=0.5]{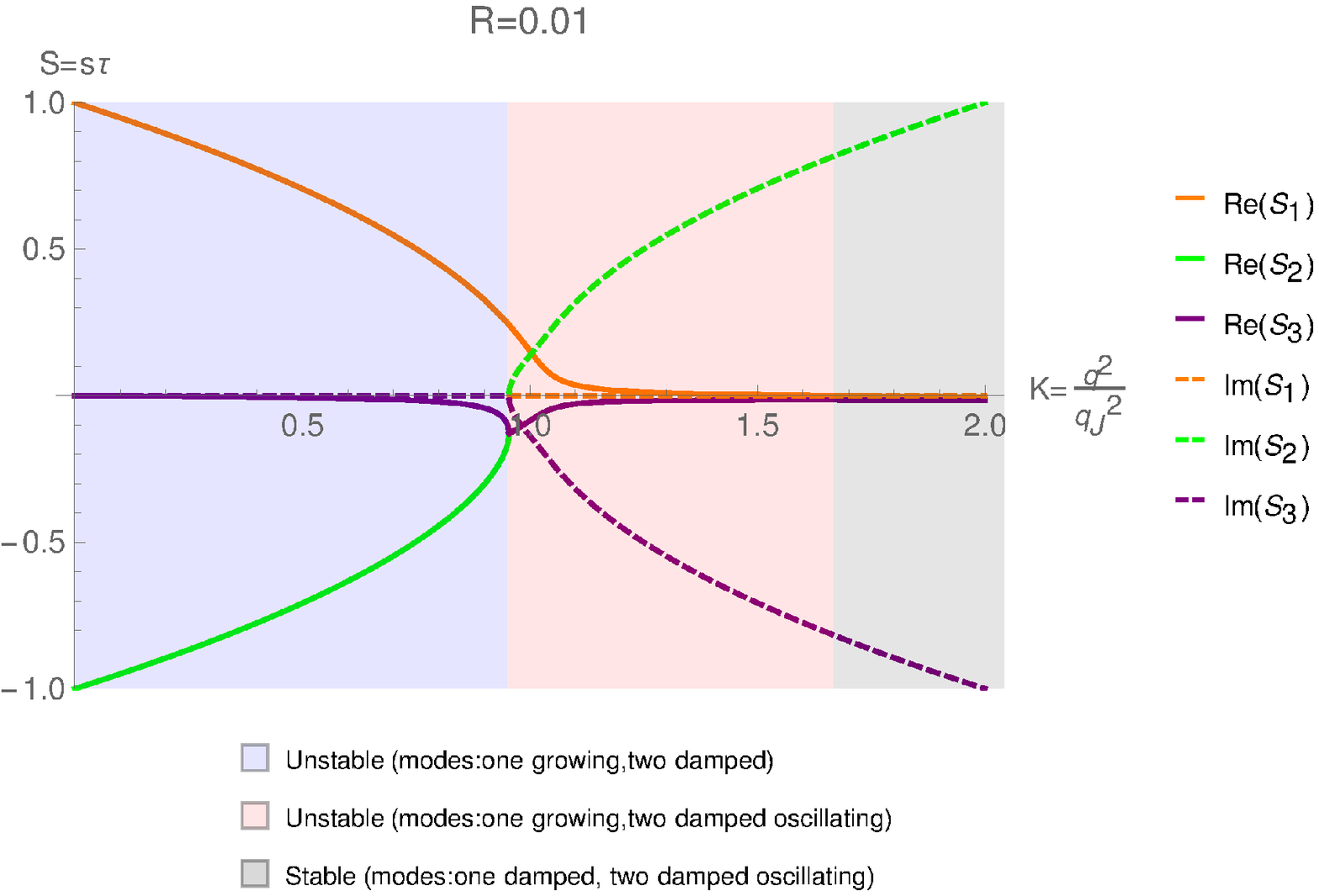}
\par\end{centering}
\caption{\label{fig:8}The real (continuous line) and imaginary (dashed line)
parts of the roots of the dispersion relation for $R=0.01$.}
\end{figure}
\par\end{center}

\begin{center}
\begin{figure}
\begin{centering}
\includegraphics[scale=0.5]{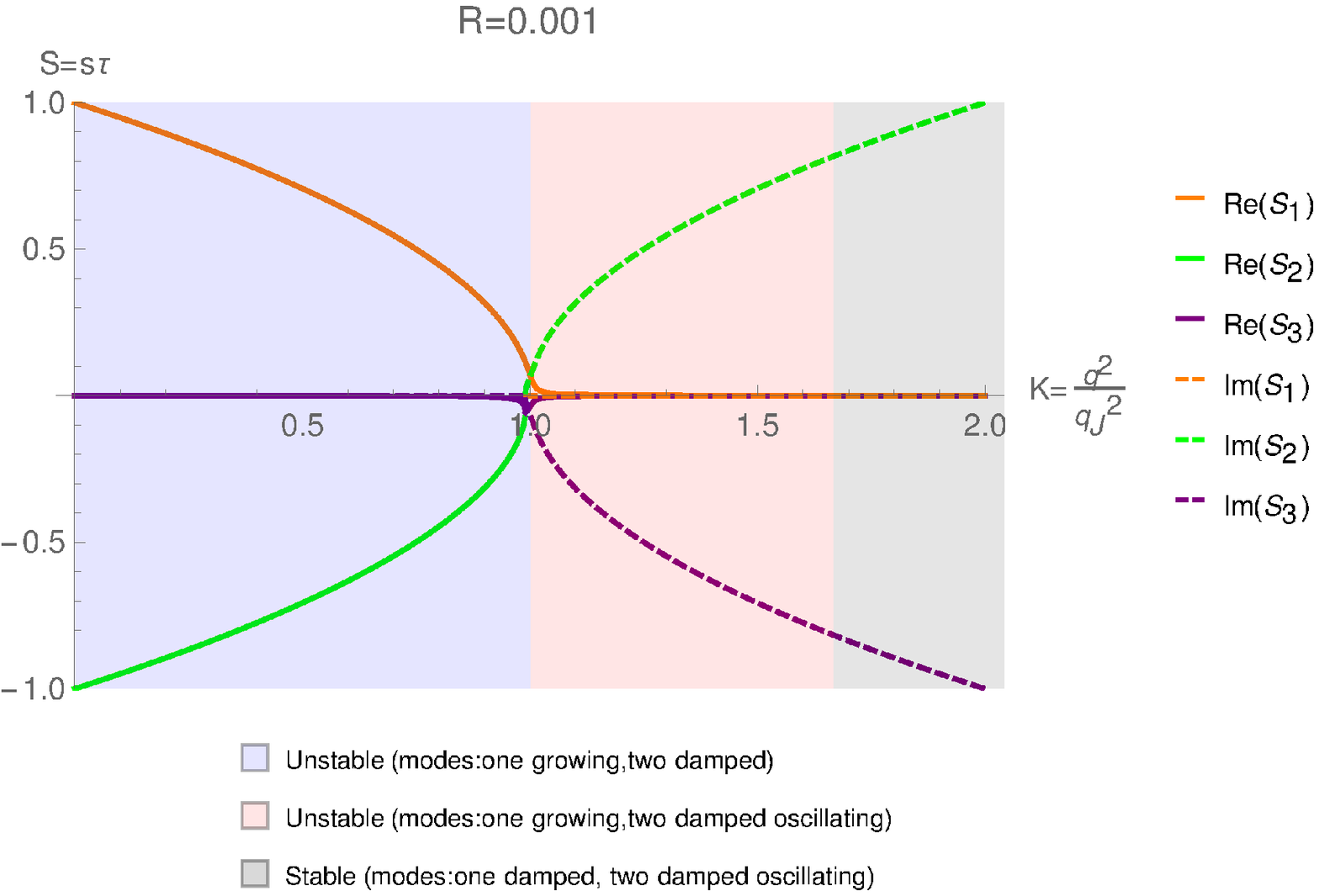}
\par\end{centering}
\caption{\label{fig:9}The real (continuous line) and imaginary (dashed line)
parts of the roots of the dispersion relation for $R=0.001$.}
\end{figure}
\par\end{center}

\noindent Figures \ref{fig:7} and \ref{fig:8} show the roots of
the discriminant for $R=0$ and $R\rightarrow0$ respectively. Qualitatively,
the behavior is similar, however, since $\Delta\left(K,R\right)\sim-4\left(K-1\right)^{3}+\mathcal{O}\left(R^{2}\right)$
the value for which the discriminant changes sign approaches one
but the real part of $S_{1}$ still only changes sign at $K=5/3$.
The similarity in the behavior observed graphically corresponds to
the fact that the real part of $S_{1}$ becomes very close to zero
abruptly near $K=1$ but remains positive until the critical value
$5/3$ is attained. 

\section{Summary and concluding remarks}

In this work, a detailed mathematical analysis of the dynamics of
density fluctuations in the linear regime for an ideal gas in the
presence of dissipation has been carried out. The gravitational potential
generated by such fluctuations can lead to structure formation for
wavenumbers below a critical value.

The analysis consisted of a detailed study of the dispersion relation
arising from the linearized set of hydrodynamic equations: mass, momentum
and internal energy balance. The critical wavenumber has been found
to differ from the usual Jean's wavenumber by a factor of $5/3$,
independently of the value of the transport coefficients as long as
they don't identically vanish and an internal energy balance equation,
featuring dissipative effects however negligible they become, is
retained.

Transitions in the behavior of the stable modes have been identified,
depending on the value of the ratio between the characteristic times
involved in the problem: gravitational and microscopic (collisional).
When dissipative effects are non-negligible,
the stable modes within the stable region are purely damped and this
behavior can change to damped oscillations, with smaller damping rates,
either within the unstable region or in the stable range of wavenumbers.
This is relevant since real roots of the dispersion relation lead
to central narrow peaks while evidence of complex values comes in
form of a doublet of broader peaks. These information could in principle
be verified in a light scattering experiment \cite{Berne}.

The calculations here presented, shed some light on the question on
whether the critical wavenumber corresponds to the usual Jeans wavenumber
or the slighter larger value obtained considering dissipation. Indeed,
the strict minimum wavelength that can lead to structure formation
is given by $2\pi\left(5q_{J}/3\right)^{-1}$. However, in the limit
of weak dissipation, the growth rate of the instability becomes negligible
when the fluctuations' wavelengths exceed $2\pi q_{J}^{-1}$. It is
important to emphasize that although fluctuations do not grow significantly
for $1\leq q/q_{J}<5/3$, they become damped only for $q>5q_{J}/3$.
These elements allow for a definitive description of the non-dissipative
instability mechanism as a limit of the dissipative one without the
need for the consideration of distinct equations of state depending
on weather an internal energy balance equation is considered in the
set of hydrodynamic equations or not. 

The problem here addressed assumes the gravitational field is weak
enough such that the classical approach is adequate. A relativistic
formalism, including a fluctuating metric can be found in Ref. \cite{Garcia-Mendez-Sandoval}
in which corrections due to dissipation, metric fluctuations and high
temperature are taken into account. A relativistic version of the
calculations presented in the present paper might be helpful in the
study of the contribution of each of the three effects mentioned for
the hot, general relativistic, dissipative ideal gas. Such a formalism
will be developed in the near future.

\section*{References}

\bibliography{ref-rel}
\bibliographystyle{unsrt}
\end{document}